\documentclass[namedreferences]{solarphysics}
\usepackage[optionalrh]{spr-sola-addons} % For Solar Physics
\usepackage{epsfig}          % For eps figures, old commands
\usepackage{graphicx}        % For eps figures, newer & more powerfull
\usepackage{color}           % For color text: \color command
\usepackage{url}             % For breaking URLs easily trough lines
\begin{document}

\begin{article}

\begin{opening}

\title{Acoustic Events in the Solar Atmosphere from \textit{Hinode}/SOT NFI observations\\ {\it Solar Physics}}

\author{J.-M.~\surname{Malherbe}$^{1}$\sep
        T.~\surname{Roudier}$^{2}$\sep
        M.~\surname{Rieutord}$^{2}$\sep
        T.~\surname{Berger}$^{3}$\sep
        Z.~\surname{Franck}$^{3}$
}

\runningauthor{J.-M. Malherbe \textit{et al.}}

\runningtitle{Acoustic Events in the solar photosphere from
\textit{Hinode}}

\institute{$^{1}$ LESIA, Observatoire de Paris, 92195 Meudon,
France,
                 email: \url{Jean-Marie.Malherbe@obspm.fr}\\
              $^{2}$ LAT, Universit\'{e} de Toulouse, CNRS, 14 Avenue Edouard Belin, 31400 Toulouse,
              France,
               email: \url{Thierry.Roudier@ast.obs-mip.fr}\\
              $^{3}$ Lockheed Martin Solar and Astrophysics Laboratory, Palo Alto, 3251 Hanover Street, CA 94303,
              USA,
               email: \url{Berger@lmsal.com}, \url{Zoe@lmsal.com}
               }

\begin{abstract}
We investigate the properties of acoustic events (AEs), defined as
spatially concentrated and short duration energy flux, in the quiet
sun using observations of a 2D field of view (FOV) with high spatial
and temporal resolution provided by the Solar Optical Telescope
(SOT) onboard \textit{Hinode}. Line profiles of Fe \textsc{i} 557.6
nm were recorded by the Narrow band Filter Imager (NFI) on a $82''
\times 82''$ FOV during 75 min with a time step of 28.75 s and
0.08$''$ pixel size. Vertical velocities were computed at three
atmospheric levels (80, 130 and 180 km) using the bisector technique
allowing the determination of energy flux in the range 3-10 mHz
using two complementary methods (Hilbert transform and Fourier power
spectra). Horizontal velocities were computed using local
correlation tracking (LCT) of continuum intensities providing
divergences.

The net energy flux is upward. In the range 3-10 mHz, a full FOV
space and time averaged flux of 2700 W m$^{-2}$ (lower layer 80-130
km) and 2000 W m$^{-2}$ (upper layer 130-180 km) is concentrated in
less than 1\% of the solar surface in the form of narrow (0.3$''$)
AE. Their total duration (including rise and decay) is of the order
of $10^{3}$ s. Inside each AE, the mean flux is $1.6~10^{5}$ W
m$^{-2}$ (lower layer) and $1.2~10^{5}$ W m$^{-2}$ (upper). Each
event carries an average energy (flux integrated over space and
time) of $2.5~10^{19}$ J (lower layer) to $1.9~10^{19}$ J (upper).
More than $10^{6}$ events could exist permanently on the Sun, with a
birth and decay rate of 3500 s$^{-1}$. Most events occur in
intergranular lanes, downward velocity regions, and areas of
converging motions.

\end{abstract}
\end{opening}
%-------------------------------------------------

\section{Introduction}

Oscillations appear all over the Sun and are the second major
component of the photospheric velocity field. 5-min oscillations
have been very closely studied in recent years particularly to probe
the physical properties of the solar interior.

Solar pressure waves (and oscillations) are believed to be excited
by turbulent convection around 200 km below the surface (Goode
\textit{et al.}, 1992), where velocities take maximum values. Recent
numerical simulations of acoustic wave propagation and dispersion
(Shelyag \textit{et al.}, 2006) suggest a possible location of the
flux source 2000 km below the solar surface.

The mechanism of excitation of oscillations is generally understood
from a theoretical perspective. Observational verifications and
investigations need to be done. The question arises whether such
excitation occurs all over the surface or whether localized sources
of enhanced generation exist.

Wave amplitude and phase diagnosis have been applied to detect
acoustic events (AEs) as sources of enhanced acoustic wave
generation at high spatial resolution (Brown, 1991). AEs (defined as
sub-arcsec spatially concentrated and short duration energy flux)
have been first detected on high-resolution spectra from ground
based observations (Rimmele \textit{et al.}, 1995; Strous \textit{et
al.}, 2000) and are now well studied by 2D spectrometers using line
scans (Bello Gonz\'{a}lez \textit{et al.}, 2009, 2010a, 2010b). AEs
(or seismic events) are visible in the photosphere but could be the
counterpart of the localized events which excite solar oscillations
beneath the photosphere, transferring mechanical energy into
oscillations. The strongest events appear to be located in the
intergranular lanes, where downflows are often detected. They may be
excited by some cooling mechanism (Rimmele \textit{et al.}, 1995).
Strous \textit{et al}. (2000) suggest that AEs could generate an
appreciable part of the energy necessary to sustain the whole
$p$-mode spectrum. More recently, Bello Gonz\'{a}lez \textit{et al.}
(2010b), using the SunRise experiment, found that the energy of
acoustic waves above 5 mHz could compensate partly energy losses of
the chromosphere and originates mainly from small and localized
regions.

Convective motions produce temporally varying stress and entropy
fluctuations which can act as sources of acoustic waves. Theoretical
work shows that downflowing plumes (cooling events; see Rast, 1999)
can locally excite acoustic oscillations. 3D numerical simulations
(Skartlien \textit{et al.}, 2000) suggest that compressed granules
at mesogranular boundaries could generate upward-propagating waves.
$p$-mode excitation is suspected to occur preferentially close to
the top of the convection zone in the superadiabatic layer, where
turbulent velocities and entropy fluctuations are the largest (Stein
\textit{et al.}, 2004).

Acoustic wave dissipation is also assumed to be an important heating
source for at least the lower chromosphere. In a recent study,
Rutten \textit{et al.} (2008) observed that acoustic grains (Ca
\textsc{ii} H2v and K2v lines) in the chromosphere could be caused
by AEs. Their photospheric diagnosis suggested that the Ca
\textsc{ii} grains could be the signature of acoustic excitation by
downdrafts. They also showed that exploding granules could explain
oscillation amplitude enhancements, which could modulate the
H$\alpha$ line core intensity above.

In this paper, we present high spatial resolution Doppler
measurements at several altitudes in the photosphere in order to
detect AEs. We take benefit of observations carried out on 4
September 2009, with complete absence of atmospheric effects and
outstanding spatial resolution in a large 2D field of view (FOV).
This provides a better statistics than before at unprecedent spatial
resolution. We investigate the location of AEs, their temporal
behaviour, dynamical and energetic properties in the range 3-10 mHz
and the relationships to the granular or intergranular pattern.

\section{Observations and Data Reduction}

We used multi-wavelength data sets of the Solar Optical Telescope
(SOT) onboard \textit{Hinode} (\textit{e.g.} Ichimoto \textit{et
al.}, 2005; Suematsu \textit{et al.}, 2008). The SOT has a 0.50 m
primary mirror aperture with a spatial resolution of about 0.28$''$
at 550 nm. Observations were performed at disk center, so that
vertical velocities produce Dopplershifts, whereas horizontal
velocities can be derived from the proper motions of granules.

We used the Narrow-band Filter Imager (NFI) to scan profiles of the
magnetically insensitive 557.6 nm Fe \textsc{i} line with a spatial
resolution of 0.28$''$. The full width at half maximum of the filter
is 60 m\AA. The spectral line was scanned using nine wavelengths
along the line profile in the following order, with respect to the
line center: -160, -120, -80, -40, 0, +40, +80, +120, +160 m\AA. The
observations were recorded continuously on 4 September 2009, from
21:07:59 to 22:23:13 UT. The solar rotation was compensated. Each
line scan took 28.75 s for a total observing time of 75 min, so that
we have 158 consecutive times of observations corresponding to 9
$\times$ 158 = 1422 images. The FOV was $82'' \times 82''$ ($1024
\times 1024$ pixels) with a pixel sampling of 0.08$''$.

We first aligned images corresponding to the blue wing of the line
at -160 m\AA~ apart from line center (blue continuum) using cross
correlation of 4 quadrants and detected the maximum of the cross
correlation functions at a fraction of pixel using 2D paraboloid
interpolation. Then we aligned together the 9 consecutive spectral
images (-160 to + 160 m\AA) covering the line and belonging to the
same scan of 28.75 s duration by correlating two successive images
spectrally separated by 40 m\AA. The procedure reduced the FOV to
$80.5''~\times~80.8''$ ($1006 \times 1010$ pixels), showing that the
tracking of \textit{Hinode} with solar rotation compensation is
excellent.

Using a narrow-band filter has the advantage of a large 2D FOV, but
the resulting counterpart is that wavelengths scanned along the line
profile are not simultaneous. In order to correct this effect, which
can induce errors in deriving radial velocities from line profiles,
we applied a phase correction to the temporal Fourier transform (FT)
of the set of data, for each wavelength, with respect to the first
wavelength of the line scan (blue continuum). After computation of
the fast Fourier transform (FFT), a phase correction of the form $
\exp(2 i \pi u \Delta t)$ was applied ($ u $ is the temporal
frequency , $ \Delta t $ with $ 0 \leq \Delta t \leq 28.75$ s is the
time lag between the blue continuum and the current wavelength), and
corrected data were obtained from the inverse FFT.

In order to improve image quality, we computed the modulation
transfer function (MTF) of the telescope taking into account the
characteristics of the primary mirror, the central occultation and
the three branches of the secondary mirror, together with the
geometry of square pixels of the CCD device. Each intensity image of
the series was deconvolved with the MTF of the telescope and the CCD
array.

We detected small transmission fluctuations in the filter (a few
percent) at constant wavelength. This effect was corrected with
respect to the mean value of the light curve calculated on the
overall FOV for each wavelength position.

\begin{figure}
\centering
\includegraphics[width=0.9\textwidth,clip=]{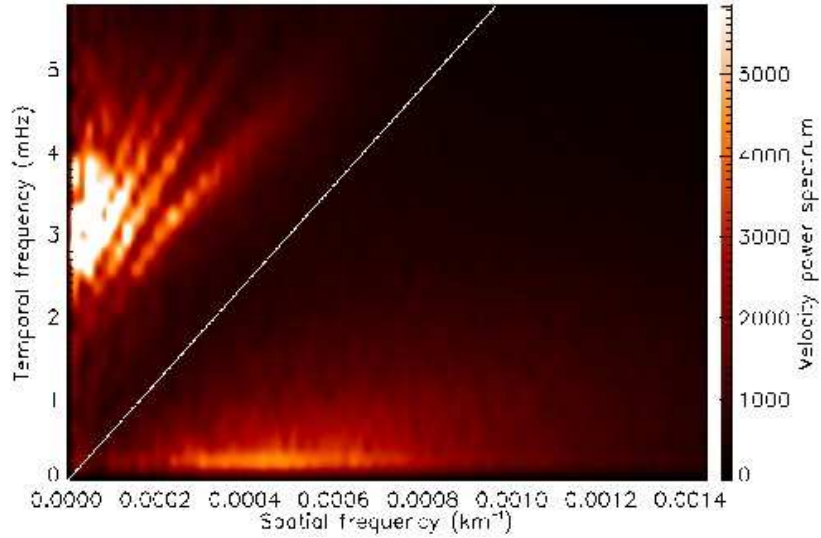}
 \caption[]{$k - \omega$ diagram of radial (vertical) velocities at the altitude of 130
 km. The abscissa and ordinate represent the spatial and temporal frequencies in km$^{-1}$ and in Hz, respectively.
 The line of $ \omega/k = C_{\textrm{s}}$ = 6 km s$^{-1}$ is also shown.} \label{spectrevC}
\end{figure}

We computed horizontal velocities from proper motions of granules
observed in the continuum (average between blue and red wings at
-160 m\AA~ and + 160 m\AA~ apart from the line centre). We first had
to remove the effects of oscillations before the detection of
granular motions; for that purpose, we applied a subsonic Fourier
filter in the $k-\omega$ space, where k and $\omega$ are
respectively the wave number and the frequency. This filter is
defined by a region of high frequencies set at zero in the
$k-\omega$ space. A 3D FFT transform was computed in the 2D space +
time and only Fourier components such that $ \omega/k\leq
C_{\textrm{s}} $ = 6 km s$^{-1}$ (where $ k =
\sqrt{k_{x}^{2}+k_{y}^{2}} $) were retained (Figure~\ref{spectrevC})
to keep only the convective contribution in the final data (Title
\textit{et al.}, 1998).

We used the local correlation tracking (LCT) as described by
November and Simon (1988) in order to compute horizontal velocities
$ v_{x} $ and $ v_{y} $ as a function of space and time, from proper
motions of the granulation in the nearby continuum of the line.
Horizontal motions were necessary to calculate the horizontal
divergence ($
\partial v_{x} /
\partial x +
\partial v_{y} / \partial y $) of the flow, as a function of
space and time.

Then we computed radial velocities and intensities at different
depths in the line profile using the bisector technique. We defined
three bisectors (chords) whose widths are 80 m\AA~ (line core), 120
m\AA~ (inflexion points) and 160 m\AA~ (line wings). The line chord
of 160 m\AA~ was used to sample a layer around an altitude of 80~km
(Altrock \textit{et al.}, 1975; Berrilli \textit{et al.}, 2002),
while chords of 120 m\AA~ and 80 m\AA, deeper in the line, were used
respectively to study atmospheric levels at about 130~km and 180~km.
For each time, a mean line profile was computed over the whole FOV,
defining a wavelength and intensity reference for the three chords
of the bisector. We applied the same method to individual profiles
of the 2D FOV and computed for the three chords Dopplershifts and
intensity fluctuations by reference to the mean profile.
Dopplershifts provided the vertical component of the velocity $
v_{z} $  as a function of space and time at three different
altitudes ($z$ = 80, 130 and 180 km, upward/downward velocities
respectively positive/negative).

\section{Temporal Behavior and Energy Flux of Acoustic Events from Hilbert Transform}

From radial velocities at different heights, we derived phases and
amplitudes of vertically travelling waves and the acoustic flux as a
function of time, spatial position in the FOV and altitude in the
photosphere.

Following  Rimmele \textit{et al.} (1995) and Strous \textit{et
al.}(2000), the acoustic flux $e$ per unit of surface (W m$^{-2}$)
can be approximated by the product of a volumic energy density and
the group velocity, assuming that the density $ \rho $ is constant:

$e$ $\propto$  $ \rho~ v_{z}^{2}~ v_{\textrm{g}}$

\noindent where $ |v_{z}| $ is the amplitude and $v_{\textrm{g}}$ is
the group velocity of the wave.

The group velocity cannot be derived directly from observations,
contrarily to the phase velocity. For sonic waves, we used the
relation $v_{\textrm{g}} v_{\varphi} = C_{\textrm{s}}^{2}$, where $
v_{\varphi} $ is the phase velocity and $ C_{\textrm{s}} =
\sqrt{\gamma P / \rho}$ is the adiabatic sound speed ($P$ pressure,
$\gamma$ = 5/3). This approximation is valid for vertically
propagating waves in a plane parallel isothermal atmosphere (no
magnetic field) with the dispersion relation: $\omega^{2} =
C_{\textrm{s}}^{2} k^{2} + N_{\textrm{s}}^{2}$. With $P$ = 4000 Pa
and $\rho = 1.23~10^{-4}$ kg m$^{-3}$ (altitude 150 km, HSRA,
Gingerich \textit{et al.}, 1971), we find $C_{\textrm{s}}$ = 7.4 km
s$^{-1}$. The cutoff frequency $N_{\textrm{s}}=\gamma g /(2
C_{\textrm{s}})$ ($g$ solar gravity) is of the order of 4.9 mHz in
this simplified model.

The phase velocity $ v_{\varphi} $ was derived from observations by:

$ v_{\varphi} = \omega / k_{z}$ with $ k_{z} =\Delta\varphi / \Delta
z $

\noindent where $ \omega $ is the frequency of the assumed quasi
monochromatic wave, $\Delta\varphi$ is the phase lag between the two
heights of observation and $ \Delta z $ is the thickness of the
layer. The group velocity is directly related to the phase lag
$\Delta\varphi$ by the relation:

$v_{\textrm{g}} = \Delta\varphi \frac{C_{\textrm{s}}^{2}}{\omega
\Delta z}$.

Numerically, for 5-min oscillations with $ \Delta z $ = 50 km,
$v_{\textrm{g}} = 0.9 ~ \Delta\varphi$ with $v_{\textrm{g}}$ in km
s$^{-1}$ and $\Delta\varphi$ in degrees (typically $|\Delta\varphi|
\leq 10\deg$ so that $|v_{\textrm{g}}| \leq C_{\textrm{s}}$).

The energy flux is given by the relation:

$e$ $\propto$  $ v_{z}^{2} (\gamma P / \omega) (\Delta\varphi /
\Delta z)$ in W m$^{-2}$.

We had to derive the velocity amplitude $|v_{z}|$ and phase lag $
\Delta\varphi $ from measurements at two different levels. We used a
mathematical method consisting to extend the observed vertical
velocity $ v_{z}(t) = v(t) $ to the complex quantity $V(t)$ such
that:

$V(t) = v(t) + i \underline{v}(t)$

\noindent where $\underline{v}(t)$ is the Hilbert transform (HT) of
$ v(t) $ (see the Appendix). We used the velocity data at two
altitudes $ z_{1}$ and $ z_{2}$ in order to build complex velocities
$V_{1}(t) = v_{1}(t) + i \underline{v}_{1}(t)$ and $V_{2}(t) =
v_{2}(t) + i \underline{v}_{2}(t)$, from which we derived the phase
lag $ \Delta\varphi = \varphi_{2} - \varphi_{1} =
\textrm{arg}(V_{2}) - \textrm{arg}(V_{1}) $ and the average
amplitude $ A = (|V_{1}| + |V_{2}|)/2$.

As the method is valid only for quasi monochromatic signals, we
filtered data to isolate temporal frequencies $u$ using the
following narrow band pass filter (full width at half maximum FWHM
of 1.2 mHz):

\begin{enumerate}
  \item $ u \leq u_{0}-1.0 $ mHz, filter null,
  \item $ u_{0}-1.0 $ mHz $\leq u \leq u_{0}-0.2 $ mHz, cos$^{2}$
  ramp,
  \item $ u_{0}-0.2 $ mHz $\leq u \leq u_{0}+0.2 $ mHz, filter at 1,
  \item $ u_{0}+0.2 $ mHz $\leq u \leq u_{0}+1.0 $ mHz, cos$^{2}$
  ramp,
  \item $ u \geq u_{0}+1.0$ mHz, filter null.
\end{enumerate}

\noindent where $ u_{0}$ is the central frequency. We have chosen
six different central frequencies: 3.3, 4.5, 5.7, 6.9, 8.1 and 9.3
mHz (for the frequency 3.3 mHz, we also used a second filter of 1.8
mHz FWHM to produce images and movies).

The phase lag and velocity group calculations were done inside two
different layers (80-130 km and 130-180 km) for which the average
pressure is respectively 5700 Pa and 4000 Pa (according to the HSRA
model atmosphere, Gingerich \textit{et al.}, 1971).

\section{Energy Flux Derived From the Fourier Power Spectra}

Following Bello Gonz\'{a}lez \textit{et al.} (2009), we used a
complementary method to compute the mean energy flux from the
Fourier power spectra (FPS) $P_{\nu}$ of velocities, defined as the
absolute value of the FT of $v^{2}(t)$. The energy flux is now
defined by:

$e$ $\propto$ $\rho~P_{\nu}~\Delta\nu~v_{\textrm{g}}$

\noindent where $\Delta\nu$ is the frequency bandwidth (we chose 1.2
mHz) and $v_{\textrm{g}}$ is the group velocity. This method is the
most accurate for determining the average energy flux (as it does
not require any frequency filtering), but on the contrary it does
not provide temporal evolution of the flux and the group velocity is
not determined from observations. We used the theoretical value
provided by the simple plane parallel model:

$v_{\textrm{g}} = C_{\textrm{s}}
\frac{\sqrt{\omega^{2}-N_{\textrm{s}}^{2}}}{\omega}$,

\noindent so that it is constant and uniform over the FOV and there
is no energy flux for frequencies below 4.9 mHz. Computations have
been performed for four frequencies (5.7, 6.9, 8.1, and 9.3 mHz) and
for three different altitudes (80, 130, and 180 km) for which the
corresponding pressure and volumic mass are respectively: 7400 Pa,
$2.1~10^{-4}$ kg m$^{-3}$; 4800 Pa, $1.45~10^{-4}$ kg m$^{-3}$; 3100
Pa, $9.8~10^{-5}$ kg m$^{-3}$ (according to the HSRA model
atmosphere, Gingerich \textit{et al.}, 1971).

Contrarily to the first method based on HT, the energy flux is now
proportional to the volumic mass instead of the pressure.

\section{Granulation, Dynamics and Acoustic Events }

We built several video movies using the 158 time steps that show the
full temporal sequence (75 min) of granulation images with AEs
(layer 130-180 km) and other quantities overlaid. Energy flux was
computed for movies at 3.3 mHz (FWHM 1.8 mHz). The movies can be
consulted in reduced size (central part of the FOV with 640 $\times$
640 pixels) in standard MPEG1 format (5 Mbytes) or in full size
(1006 $\times$ 1010 pixels) as MPEG4 (20 Mbytes) or non compressed
JavaScript (40 Mbytes) at:

\url{http://www.lesia.obspm.fr/perso/jean-marie-malherbe/Hinode2010/index.html}.

Movie 0 summarizes the sequence of raw data processed here.

\begin{figure}
\centering
\includegraphics[width=0.9\textwidth,clip=]{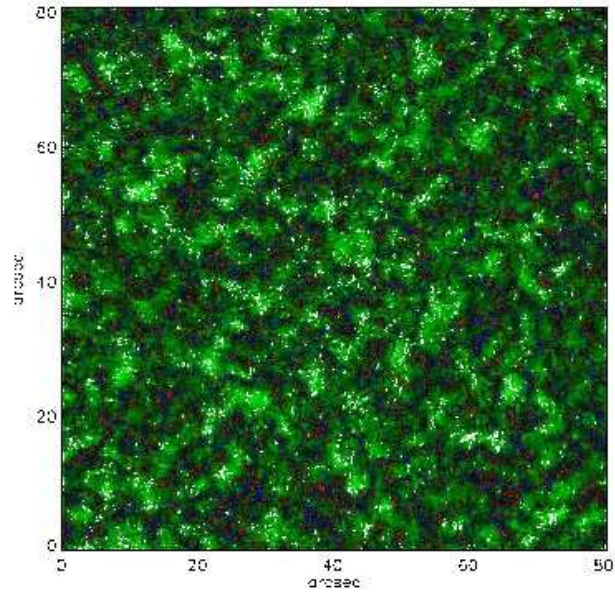}
 \caption[]{Acoustic flux (3.3 mHz, 1.8 mHz FWHM) between 130 and 180 km from chords 80 and
120 m\AA, amplitude and phase lag. AEs are indicated by white dots.
 Green color represents velocity amplitude, and blue and red indicate positive and negative phase lags,
  respectively. FOV = $80''\times80''$.} \label{amphase}
\end{figure}

Figure~\ref{amphase} (and movie 1) shows the unsigned vertical
acoustic flux in the middle of the sequence (image 76). The flux is
spatially concentrated over a few pixels (1 pixel = 0.08$''$). AEs
(flux proportional to the square of velocity amplitudes and phase
lags) are characterized by high velocity amplitudes combined to
moderate phase lags between both levels of observation (typically a
few $\deg$ or a few km s$^{-1}$ for the group velocity). In regions
of low amplitude, phase lags become noisy (this is a limitation of
the HT), generating small but noisy flux. The mean temporal
evolution of a typical AE is detailed in Section 6.

\begin{figure}
\centering
\includegraphics[width=0.9\textwidth,clip=]{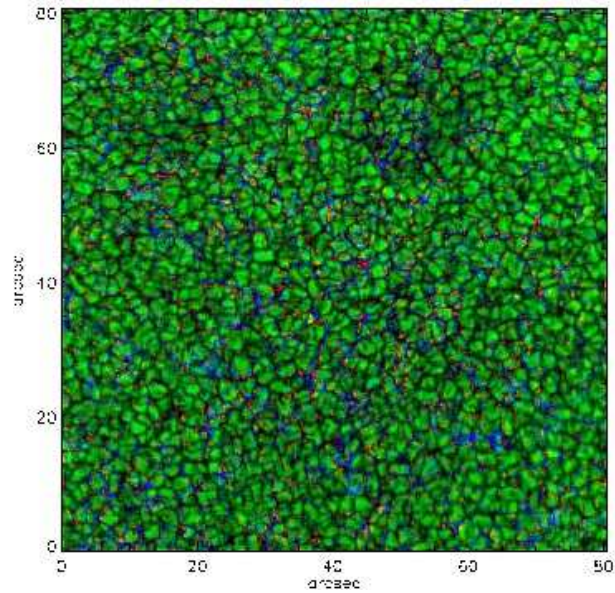}
\caption[]{Acoustic flux (3.3 mHz, 1.8 mHz FWHM) between 130 and 180
km from chords 80 and 120 m\AA. Green color represents continuum
intensity, and blue and red points indicate upward and downward
concentrations of acoustic flux, respectively. FOV =
$80''\times80''$.} \label{fluxac}
\end{figure}

Figure~\ref{fluxac} (and movie 2) shows the signed vertical acoustic
flux in the middle of the sequence with respect to the granulation
pattern observed in the continuum, defined as the average of the
wing intensities at -160 m\AA~ and +160 m\AA. Continuum images were
generated after removal of high frequencies in the $ (k - \omega) $
diagram. In the $ (k_{x}, k_{y}, \omega) $-space, we filtered all
frequencies (oscillations) corresponding to $ \omega \geq
C_{\textrm{s}} k$, where $C_{\textrm{s}}$ = 6 km s$^{-1}$ is the
sound speed, and $ k = \sqrt{k_{x}^{2}+k_{y}^{2}}$, in order to keep
only permanent convective flows. This figure, together with the
movie, shows that AEs are spatially concentrated in intergranular
lanes, and that upward propagating flux is dominant.

\begin{figure}
\centering
\includegraphics[width=0.9\textwidth,clip=]{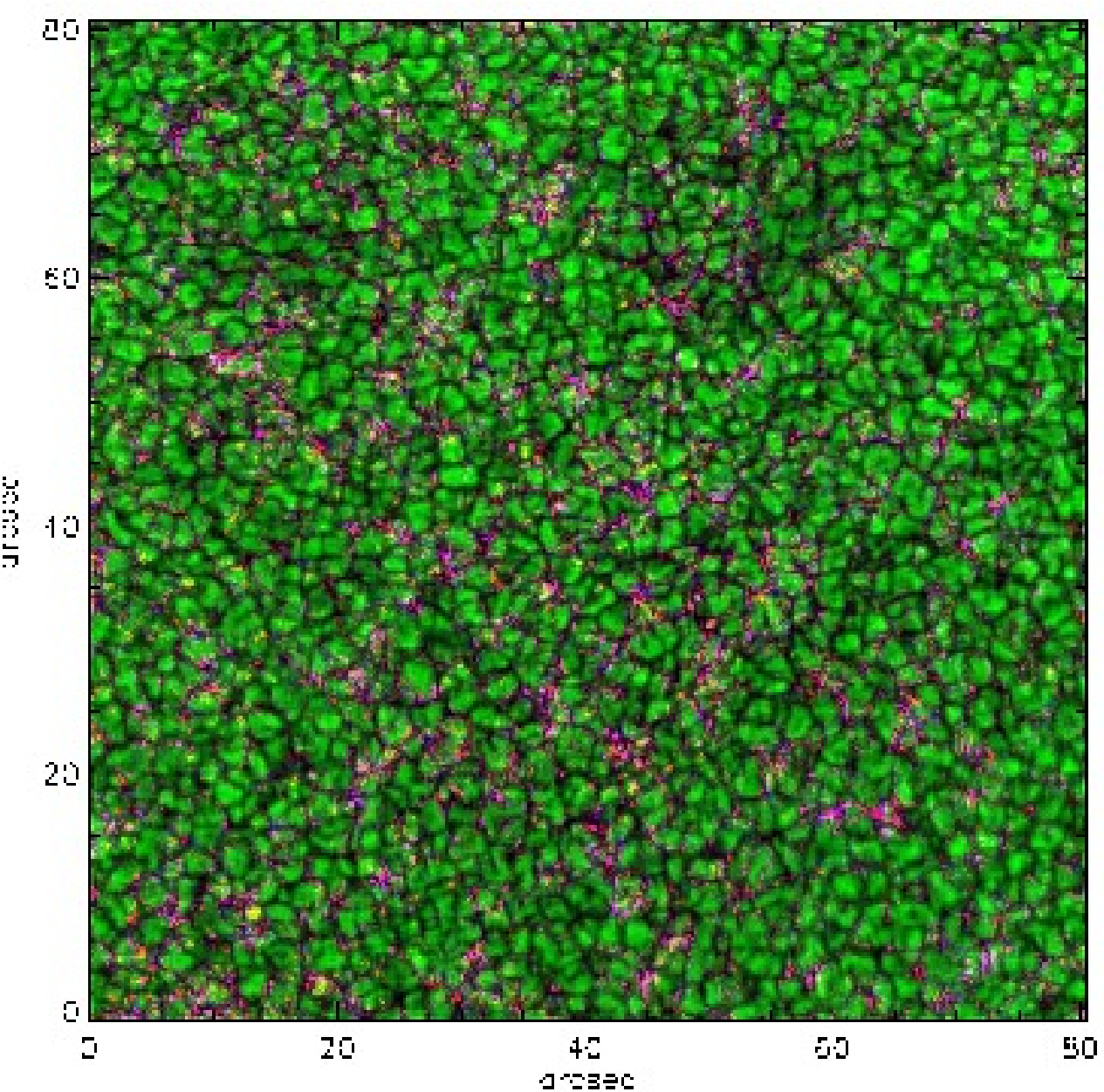}
 \caption[]{Unsigned acoustic flux (3.3 mHz, 1.8 mHz FWHM) in two different layers
with respect to the granulation pattern. Green color represents
continuum intensity, and blue and red points indicate acoustic flux
in height ranges 80 - 130 km and 130 - 180 km, respectively. FOV =
$80''\times80''$.} \label{flux2niv}
\end{figure}

The acoustic flux was computed for layers 80-130 km and 130-180 km.
Figure~\ref{flux2niv} (and movie 3) shows the unsigned flux in the
two different layers (middle of the sequence) with respect to the
granulation pattern observed in the continuum, defined as the
average of the wing intensities and suppressing high frequencies in
the $ (k - \omega) $ diagram to eliminate oscillations. The flux in
the two different layers, when perfectly superimposed, appears as
magenta; we notice that this is the case of most AEs, suggesting a
good correlation (numerically 0.83) between both layers and
propagation along 100 km for 97.3 \% of observed events.

\begin{figure}
\centering
\includegraphics[width=0.9\textwidth,clip=]{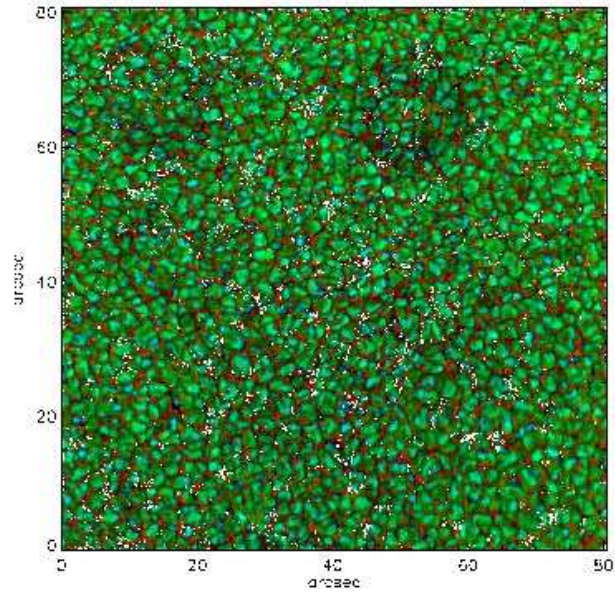}
 \caption[]{Vertical convective velocities and AEs  with
respect to the granulation pattern. AEs are indicated by white dots.
Green color represents continuum intensity, and blue and red
indicate upward and downward velocities, respectively. FOV =
$80''\times80''$.} \label{icontv}
\end{figure}

We studied the location of AEs not only with respect to the
granulation pattern, but also with respect to vertical motions. For
that purpose,
 Figure~\ref{icontv} (see also
movie 4) shows the convective vertical velocities at 130 km together
with AEs (layer 130-180 km) in the middle of the sequence
superimposed to the granulation pattern observed in the continuum.
Both vertical velocities and continuum intensities were suppressed
at high frequencies ($ \omega \geq C_{\textrm{s}} k$, where
$C_{\textrm{s}}$ = 6 km s$^{-1}$). The sequence confirms that
intergranular lanes are associated with downflows while granules are
associated to upflows. It also shows that AEs mostly occur in
intergranular lanes together with downdrafts.

\begin{figure}
\centering
\includegraphics[width=0.9\textwidth,clip=]{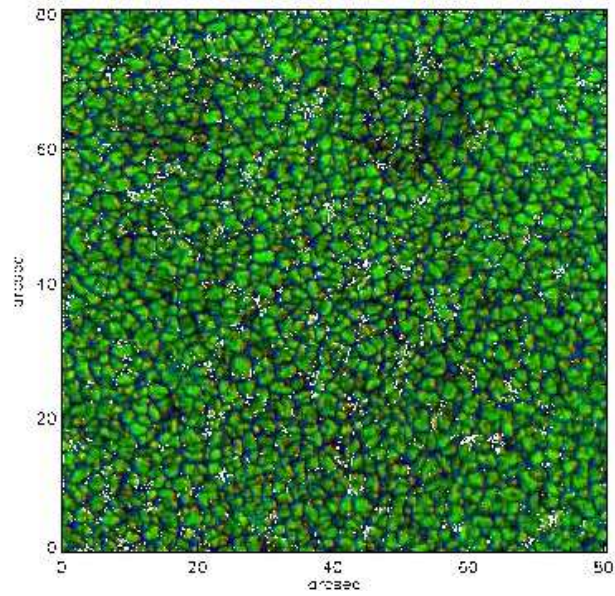}
 \caption[]{Vertical gradient of convective velocities and AEs
   with respect to the granulation pattern. AEs are indicated by white
   dots. Green color represents continuum intensity,
and blue and red indicate increasing and decreasing vertical
velocities with height, respectively. FOV = $80''\times80''$.}
\label{gradientv}
\end{figure}

Our multilevel velocity data allowed us to analyze the vertical
velocity gradient, defined as the difference of radial velocities
(positive upwards) between altitudes of 180 and 130 km. A positive
gradient means increasing (signed) velocities with height.
 Figure~\ref{gradientv} (and movie 5) shows the vertical gradient of
convective velocities together with AEs in the middle of the
sequence with respect to the granulation pattern. The gradient is
mostly positive (which means increasing velocity with height) in
intergranular lanes. In absolute value, it means that the velocity
decreases with height above intergranules. On the contrary, the
gradient is almost zero or slightly negative above granules.

Histograms (not shown) of the continuum intensity as a function of
vertical velocities or velocity gradients computed over the whole
FOV and over the total duration of the sequence confirmed a strong
correlation between dark regions (intergranules) and downdrafts, as
well as bright regions (granules) and upward velocities. Flows are
in the range -1 km s$^{-1}$ to +1 km s$^{-1}$. Dark lanes
(intergranules) are also related to positive velocity gradients (+
0.2 km s$^{-1}$ typical).

\section{Mean Properties and Temporal Evolution of Acoustic Events and Energy Flux}

\subsection{Energy Distribution of Acoustic Flux}

\begin{figure}
\centering
\includegraphics[width=0.9\textwidth,clip=]{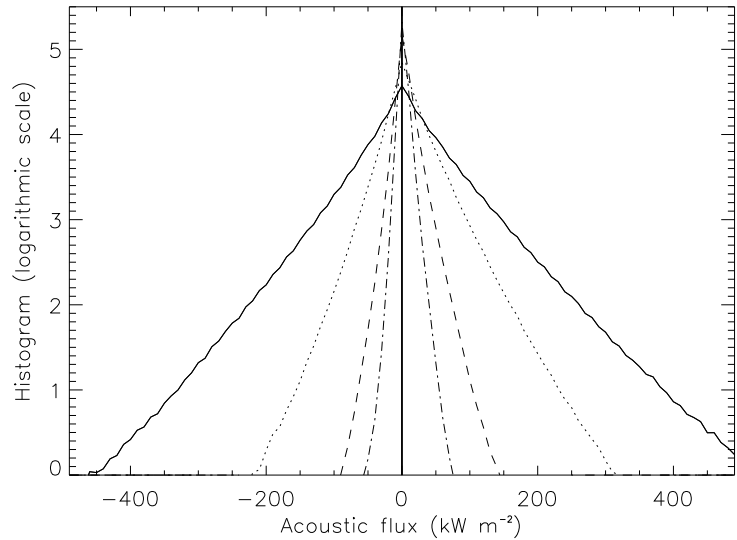}
 \caption[]{Histogram of acoustic flux between 130 and 180 km (logarithmic scale)
over the whole FOV for 3.3 mHz (solid line), 4.5 mHz (dotted line),
5.7 mHz (dashed line) and 6.9 mHz (dash-dot). The fluxes are summed
over a FWHM of 1.2 mHz.} \label{courbea}
\end{figure}

The distribution of seismic flux derived from the full FOV is
displayed in Figure~\ref{courbea}. It is qualitatively comparable at
3.3 mHz to the one got by Strous \textit{et al.} (2000), except that
this is not the same spectral line (different altitude) and
bandwidth. In the HT method, the acoustic flux is sensitive to a
multiplicative parameter, the pressure. We have taken 4000 Pa, which
represents a realistic order of magnitude for the layer 130-180 km
according to Stebbins and Goode (1987) and from the HSRA model. It
is also inversely sensitive to the thickness of the layer (50 km).
The flux distribution is asymmetric, as upward flux dominates.

\subsection{Number of Acoustic Events and Their Birth Rate}

We labeled all AEs (detected at 3.3 mHz) with flux averaged over the
whole temporal sequence above $3 \sigma$ (75 kW m$^{-2}$). This
spatial criterion allows to select energetic AEs covering less than
1\% of the surface. We used the binarization technique described by
Roudier \textit{et al.} (2003) to count and recognize pixels
belonging to the same event. We found that the number of AEs is
almost constant as a function of time within our $80'' \times 80''$
FOV: 1250 events permanently above $3 \sigma$, corresponding to
$2.2~10^{6}$ events on the whole surface of the Sun. With a typical
lifetime of 600 s at half maximum flux, we found a birth rate of
3500 s$^{-1}$ over the full Sun, or $5.7~10^{-16}$ s$^{-1}$
m$^{-2}$. This result is not far from the one ($8~10^{-16}$ s$^{-1}$
m$^{-2}$) obtained by Strous \textit{et al.} (2000).

\subsection{Energy Flux of Acoustic Events in Granules and Intergranules}

We investigated the temporal behaviour of the total acoustic flux
measured in AEs at 3.3 mHz, as a function of time, in intergranules
(negative continuum) as well as in granules (positive continuum).
AEs were selected in terms of energy flux above $3 \sigma$ (covering
less than 1\% of the surface). The net total flux of AEs (downward +
upward) dominates in intergranules (four times larger than in
granules), is upward and remains approximately constant in time,
indicating that there exists a permanent renewal of a large number
of events during the sequence.

\subsection{Power and Energy of Acoustic Events}

We have computed, using the HT, the net acoustic flux (balance
between upward and downward flux) in the range 3-10 mHz (1.2 mHz
FWHM), layers 130-180 km and 80-130 km, as a function of the
strength of AEs. We considered them as significant when the time
averaged energy flux over the whole sequence duration is spatially
above $3 \sigma$ or 75 kW m$^{-2}$, to cover less than 1\% of the
surface. In the upper layer, the full FOV averaged flux (above $3
\sigma$) was 2000 W m$^{-2}$ (2700 W m$^{-2}$ in the lower layer).
But the flux averaged only over the area of AEs in the upper layer
was $1.2~10^{5}$ W m$^{-2}$ in the range 3-10 mHz and $2.8~10^{4}$ W
m$^{-2}$ in 5-10 mHz. In the lower layer, we got $1.6~10^{5}$ W
m$^{-2}$ in the interval 3-10 mHz and $3.6~10^{4}$ W m$^{-2}$ in
5-10 mHz. The mean energy (flux integrated over space and time)
carried out by each AE in the upper layer was $1.9~10^{19}$ J in
3-10 mHz and $4~10^{18}$ J in 5-10 mHz. These values are of the same
order of magnitude than those found by Strous \textit{et al.}
(2000). In the lower layer, we got $2.5~10^{19}$ J for 3-10 mHz and
$5~10^{18}$ J for 5-10 mHz. The mean area of AE is about 12 pixels
of 0.08$''$ (corresponding approximately to the resolving power of
the telescope).

\begin{figure}
\centering
\includegraphics[width=0.9\textwidth,clip=]{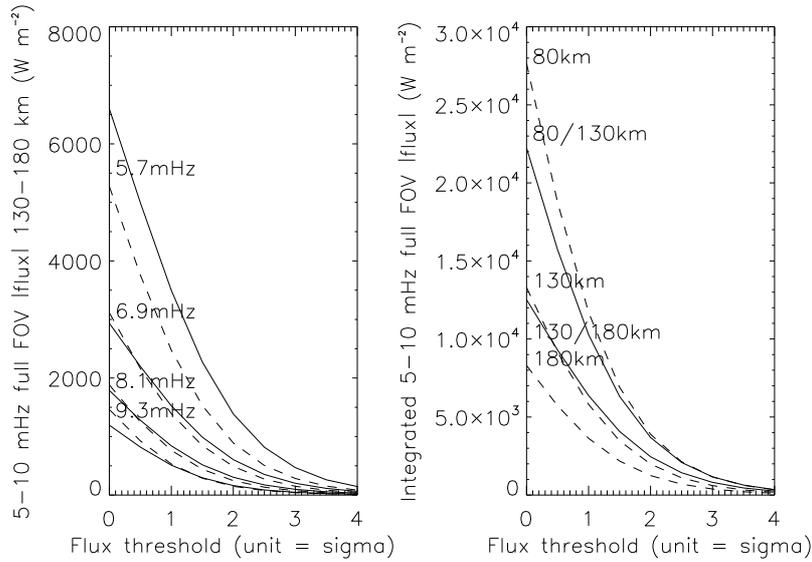}
 \caption[]{Left: Unsigned acoustic flux (W m$^{-2}$) in the 130-180 km layer averaged in time and space
 for frequencies 5.7, 6.9, 8.1, and 9.3 mHz. The solid and the dashed lines indicate results from
 HT and FPS, respectively.
 Right: Integrated 5-10 mHz unsigned flux (W m$^{-2}$)
 for layers 130-180 and 80-130 km for HT (solid
 lines). Similarly the dashed lines indicate the results from FPS at altitudes 80, 130, and 180
 km. The flux above $2 \sigma$, $3 \sigma$ and $4 \sigma$ cover
 6\%, 1\%, and 0.15\% of the solar surface, respectively.}
 \label{flux2}
\end{figure}

Figure~\ref{flux2} allows to compare results in the 5-10 mHz range
over the full FOV obtained from the HT (allowing the determination
of time dependent group velocity and signed flux inside layers
130-180 and 80-130 km) and those (unsigned) got from classical FPS
with uniform and constant group velocity
($v_{\textrm{g}}=C_{\textrm{s}}\frac{\sqrt{\omega^{2}-N_{s}^{2}}}{\omega}
$) at three altitudes (80, 130, 180 km). In order to facilitate the
comparison, we have taken the absolute value of the flux provided by
the HT; and we computed the average value of the flux between
130-180 km (upper layer) and 80-130 km (lower layer) of the FPS. The
plot for four frequencies (5.7, 6.9, 8.1, 9.3 mHz, each band having
1.2 mHz FWHM) shows that there is a good agreement between space and
time averaged results derived from the two approaches. This result
indicates that the group velocity, derived from observations by the
HT, is close in average to the one used in the FPS, emphasizing the
stochastic character of the large number of acoustic events.

We also computed (Figure~\ref{flux2}) the flux integrated over the
frequency band of 5-10 mHz inside the upper and lower layers (HT),
and at 80, 130 and 180 km (FPS). Here again, the agreement is good.
We find for the three altitudes respectively 27000 W m$^{-2}$, 12000
W m$^{-2}$ and 8000 W m$^{-2}$ over the full FOV (without any flux
threshold). Bello Gonz\'{a}lez \textit{et al.} (2009) found 2000 W
m$^{-2}$ in the same spectral line at 250 km and in Fe \textsc{i}
543.4 nm at 500 km (Bello Gonz\'{a}lez \textit{et al.}, 2010a); they
also got results (Bello Gonz\'{a}lez \textit{et al.}, 2010b) with
SunRise (no atmospheric effects) in the range 7000 W m$^{-2}$
(wavelet) to 12000 W m$^{-2}$ (Fourier) in Fe \textsc{i} 525.02 nm
at 250 km which could be compatible with our results (8000 W
m$^{-2}$ at 180 km). Concerning the contribution of AEs to the total
flux over the whole FOV, we found that events selected above $3
\sigma$ (1\% of the surface) have a contribution of less than 10\%
in the 5-10 mHz interval.

\subsection{Temporal Evolution of Acoustic Events - Amplitude, Phase, and Divergence}

AEs are energetic phenomena concentrated in space and time. Thanks
to the HT, it is possible to study the temporal evolution and the
sign of the energy flux for specific frequencies and heights.

\begin{figure}
\centering
\includegraphics[width=0.9\textwidth,clip=]{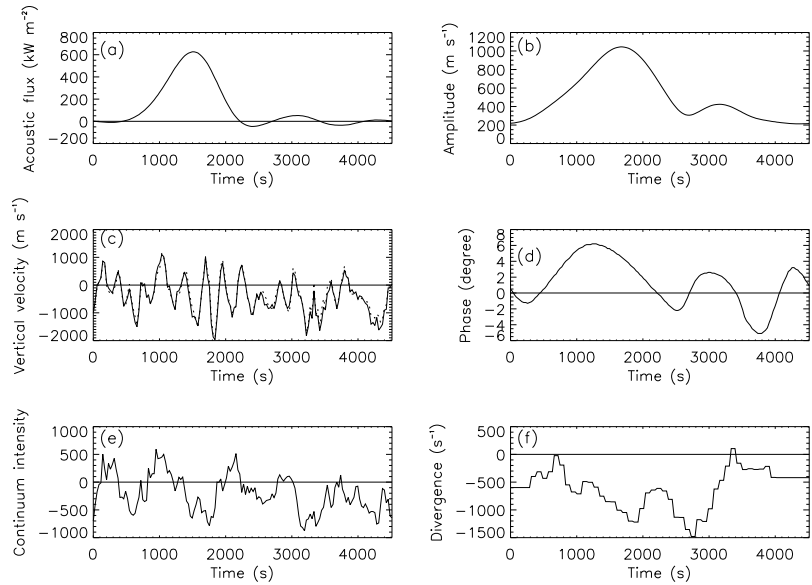}
 \caption[]{Time evolution of a typical AE in the layer 130-180 km.
  (a) Acoustic flux (kW
m$^{-2}$) at 3.3 mHz (FWHM 1.2 mHz); (b) velocity amplitude (m
s$^{-1}$);
  (c) vertical velocity (m s$^{-1}$) at 130 and 180 km; (d) phase lag ($\deg$);
  (e) continuum
intensity (arbitrary unit, dark is negative);
 (f) horizontal velocity divergence (s$^{-1}$, negative for converging flows).}
\label{evenement}
\end{figure}

First, Figure~\ref{evenement} shows the temporal behaviour of a
typical AE during the full duration of the sequence (75 min).

\begin{figure}
\centering
\includegraphics[width=0.9\textwidth,clip=]{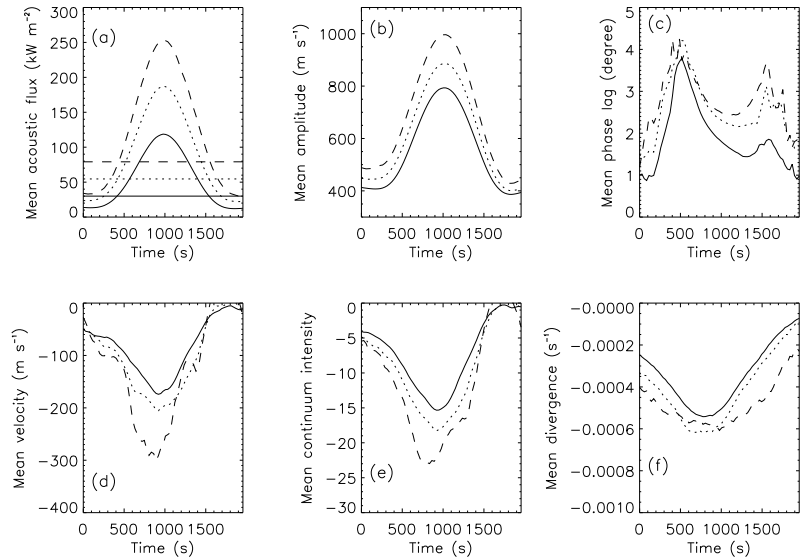}
 \caption[]{Properties of AEs averaged in space in the layer 130-180 km
 as a function of time.
Solid, dotted, and dashed lines indicate weak, medium, and intense
AEs with flux above $2 \sigma$, $3 \sigma$ and $4 \sigma$ levels,
respectively. The statistics is strongly reduced for intense events
(4 $\sigma$). (a) Acoustic flux (kW m$^{-2}$) at 3.3 mHz (FWHM 1.2
mHz); (b) velocity amplitude (m s$^{-1}$);
 (c) phase lag ($\deg$); (d) convective velocity (m s$^{-1}$); (e) continuum
intensity (arbitrary unit, dark is negative);
 (f) divergence (s$^{-1}$, negative for converging flows).}
\label{even_moyen}
\end{figure}

Then, we isolated all AEs of the sequence and analyzed their average
properties and in particular their mean temporal behaviour in the
130-180 km layer, according to the strength of acoustic flux. For
that purpose, we used different spatial thresholds in terms of time
averaged energy flux over the whole sequence duration at $2 \sigma$,
$3 \sigma$, and $4 \sigma$ (respectively 50, 75, and 100 kW
m$^{-2}$). This criterion selects events covering respectively 6\%,
1\%, and 0.15\% of the surface. We shifted in time events belonging
to these three classes after detection of their maxima, superimposed
them peak to peak, and made an average. The result is shown in
Figure \ref{even_moyen}. The mean evolution time of AEs measured at
half maximum lies around 600 s with a full duration up to 1200 s.
There is a time lag between maximum amplitudes and phases of about
500 s (300 s for Strous \textit{et al.}, 2000). Phase lags lie
around 3 to 4 degrees (corresponding to group velocities of 3 to 4
km s$^{-1}$) and velocity amplitudes are in the range 0.8 to 1.0 km
s$^{-1}$ at maximum. The strength of the flux is strongly related to
velocity amplitudes. Acoustic flux occurs essentially in
intergranules, with downward motions (-0.2 km s$^{-1}$) and negative
continuum intensity (dark regions); both quantities are correlated
(darkest lanes and fastest downflows) with the flux strength. AEs
appear also related to regions of converging flows, but the
convergence (negative divergence) seems to be maximum 200 s prior to
the flux maximum. Velocity amplitudes, downward flows, convergences
and intergranular darkness are the largest for the most intense AEs.

\section {Discussion and Conclusions}

AEs, defined as time averaged energy flux over the whole sequence
duration above $3 \sigma$ or 75 kW m$^{-2}$ (spatial criterion
selecting less than 1\% of the surface), were evidenced from the
SOT/NFI onboard \textit{Hinode} with outstanding spatial resolution
and FOV using the non magnetic Fe \textsc{i} 557.6 nm line.
Velocities were computed at three altitudes corresponding to
different chords in the line profiles, using the bisector technique.
We used two methods to determine acoustic flux. The first one, based
on the HT, after temporal frequency filtering, allowed us to
determine velocity amplitudes and phase lags as a function of time
in two layers (130-180 and 80-130 km). The group velocity is
proportional to phase lags and was used to derive the signed energy
flux. The second method (avoiding filtering) was based on FPS at
three altitudes (80, 130, and 180 km) using a constant and uniform
group velocity taken from models, but did not provide flux sign nor
temporal evolution. Both methods delivered complementary results in
agreement in the range 5-10 mHz.

In the 3-10 mHz range, the contribution of AEs is 2000 W m$^{-2}$
(upper layer 130-180 km) and 2700 W m$^{-2}$ (lower layer 80-130
km), with more flux in the 3-5 mHz than in the 5-10 mHz domain. AEs
are spatially concentrated (0.3$''$ typical, some are probably below
the resolving power of the telescope). Their average duration is
about 600 s at half maximum (1200 s including rise and decay phases)
and they carry energy mainly upwards. Each event carries in the
range 3-10 mHz an average flux of $1.2~10^{5}$ W m$^{-2}$ (upper
layer) and $1.6~10^{5}$ W m$^{-2}$ (lower layer), and a total energy
of about $1.9~10^{19}$ J (upper layer) or $2.5~10^{19}$ J (lower
layer). More than $10^{6}$ events exist at any instant on the Sun,
with a mean birth rate of 3500 s$^{-1}$. Most occur in intergranular
lanes, downward velocity regions, and areas of converging motions.
The peak of velocity amplitude (0.8 to 1 km s$^{-1}$) occurs after
the maximum phase lag, which indicates an energy propagation speed
of several km s$^{-1}$. Just before the time of peak flux,
converging flows seem to intensify. The strength of AEs is
correlated to the strength of convective flows, converging
horizontal motions, and also to the darkness of intergranules.
Regions with fastest downward motions and strongest convergences, as
well as darkest structures, produce the most energetic AEs.

From a theoretical point of view, the details of the excitation
mechanism are suspected to be stress and entropy fluctuations.
Analysis of observations and numerical simulations show that the
sources of the solar oscillations are associated with downdrafts in
dark intergranular lanes (Rimmele \textit{et al.}, 1995), and near
the boundaries of granules (Stein and Nordlund, 2001), as found
here. It will be of great interest to locate AEs with respect to
families of granules which are formed from the evolution of trees of
fragmenting granules, and which are related to larger scales as the
mesogranulation and supergranulation (Roudier \textit{et al.},
2003). This aim can be reached using a much longer observing run (6
h) than the present one. Such a run was performed (14 April 2010)
and results will be presented in a further paper.

%%%%%%%%%%%%%%%%%%%%%%%%%%%%%%%%%%%%%%%%%%%%%%%%%%%%%%%%%%%%%%%%%%%%%%%%%%%
\begin{acks}

We are grateful to the anonymous referee for helpful comments and
suggestions and to R. Stein for useful hints. We are also indebted
to the \textit{Hinode} team for the possibility to use their data.
\textit{Hinode} is a Japanese mission developed and launched by
ISAS/JAXA, collaborating with NAOJ as a domestic partner, NASA and
STFC (UK) as international partners. Scientific operation of the
\textit{Hinode} mission is conducted by the \textit{Hinode} science
team organized at ISAS/JAXA. This team mainly consists of scientists
from institutes in the partner countries. Support for the
post-launch operation is provided by JAXA and NAOJ (Japan), STFC
(U.K.), NASA, ESA, and NSC (Norway).

This work was supported by the Centre National de la Recherche
Scientifique (C.N.R.S., UMR 8109 and 5572) and by the Programme
National Soleil Terre (P.N.S.T.), France.

\end{acks}

%%%%%%%%%%%%%%%%%%%%%%%%%%%%%%%%%%%%%%%%%%%%%%%%%%%%%%%%%%%%%%%%%%%%%%%%%%%
%% Appendix
%
\appendix

In order to compute the complex signal $V(t) = v(t) + i
\underline{v}(t)$, we used the temporal FFT together with the
following properties of the Hilbert transform (HT). Given an
observed signal $v(t)$, the HT $\underline{v}(t)$ is given by:

$\underline{v}(t) = h(t) \star v(t)$

\noindent where $ \star $ denotes the convolution product and $ h(t)
= \frac{1}{\pi t}$.

For a sinusoidal signal of the form $ v(t) \propto
\textrm{cos}(\omega t + \varphi) $, the HT is simply
$\underline{v}(t) \propto \textrm{sin}(\omega t + \varphi)$, so that
in this case $V(t) \propto \exp[i (\omega t + \varphi)]$.

The easiest way to get $\underline{v}(t)$ is to use the Fourier
transform (FT with $u$ denoting the temporal frequency):

$ \textrm{FT}_{\underline{v}}(u) = \textrm{FT}_{h}(u)
\textrm{FT}_{v}(u)$

It can easily be shown that

$ \textrm{FT}_{h}(u)$ = - i $\textrm{sgn}(u)$

\noindent where $ \textrm{sgn}(u) $ is the sign function, +1 for $ u
> 0 $, 0 for $ u = 0 $ and -1 for $ u < 0 $.

As a consequence, the FT of the complex signal $V(t) = v(t) + i
\underline{v}(t)$ is given by:

$ \textrm{FT}_{V}(u) = \textrm{FT}_{v}(u)$ + i $\textrm{FT}_{h}(u)
\textrm{FT}_{v}(u) = \textrm{FT}_{v}(u) ( 1 + \textrm{sgn}(u) ) $.

This expression reduces simply to:

\begin{enumerate}
  \item $ u < 0, \textrm{FT}_{V}(u) = 0 $,
  \item $ u = 0, \textrm{FT}_{V}(u) = \textrm{FT}_{v}(u) $,
  \item $ u > 0, \textrm{FT}_{V}(u)$ = 2 $\textrm{FT}_{v}(u) $.
\end{enumerate}

We first computed the temporal FT of the observed signal $v(t)$; in
order to get the FT of the complex velocity $V(t) = v(t) + i
\underline{v}(t)$, values corresponding to negative frequencies were
eliminated and values corresponding to positive frequencies were
doubled. Finally, the complex velocity $V(t) = v(t) + i
\underline{v}(t)$ was derived from the inverse FT.

%%% BIBLIOGRAPHY %%%%%%%%%%%%%%%%%%%%%%%%%%%%%%%%%%%%%%%%%%%%%%%%%%%%%%%%%%%

\end{article}

\end{document}